# Multiscale modeling for effects of elastic substrates on three-dimensional thin-film lubrication

Z.-B. Wu[1],2

[1]LNM, Institute of Mechanics, Chinese Academy of Sciences, Beijing 100190, China

[2]School of Engineering Science, University of Chinese Academy of Sciences, Beijing 100049, China

April 30, 2019

---

[1]Email address: wuzb@lnm.imech.ac.cn

**Abstract**

For three-dimensional mono-layer molecularly thin-film lubrication, it is found that elasticity of the substrate affects tribological behaviors of a thin fluid film confined by two solid substrates. A critical thickness of the elastic substrate is determined to keep a hysteresis loop in the forward and backward processes as a signature of the qualitative tribological properties of the thin-film lubrication with the elastic substrate. To account for the elastic effects, a multiscale method that combines an atomistic description of the near region with a coarse-grained description of the far region of the solid substrate is developed to simulate the thin-film lubrication. It is shown that in a range of temperature and film-substrate coupling strength, the multiscale method yields results in excellently agreement with those obtained from the fully atomistic simulation, which reveals that the elastic response of substrate can be effectively rendered in the hybrid scheme. In the application of the multiscale method to investigate tribological properties of the multi-layer molecularly thin-film lubrication, it is found that the systematic static friction coefficient monotonously decreases as the molecular layer number in the fluid film increases. In comparison with the mono-layer molecularly thin-film lubrication, the multi-layer molecularly thin-film lubrication plays the functions of reducing friction and wear of the system by decreasing the systematic static friction coefficient.

# 1 Introduction

Nanotribology, which describes the behaviors and mechanisms of friction, wear and lubrication in relatively moving surfaces at nanoscale, is one of frontier subjects of tribology. As a prototype nanotribological system, thin-film lubrication, which consists of two solid substrates separated by a thin fluid film, plays a fundamental role in the development of advanced ultra-precision mechanical equipment and micro-machine[1, 2]. On the macroscopic scale, the surfaces of the solid substrates appear perfectly smooth. The thin fluid film can protect the surfaces from being damaged when the surfaces are sheared, and can also reduce the friction force or friction coefficient in the dry friction. However, on the microscopic scale, the surfaces actually make molecular contacts at relatively few microscopic asperities. The fluid molecules may diffuse in the interface, as well as be adsorbed and adhered to the solid surfaces. From the macroscale to nanoscale, the solid substrates exhibit multiple spatial scales and influence on the microscopic motions of both the solid and fluid molecules near or within the film. To gain deeper insights into the thin-film lubrication, it is necessary to know both the molecular motions at the interface and the effects of elastic deformation in the extended regions of the solid substrates[3, 4, 5, 6, 7, 8].

Rapid advances in computer technology are revolutionizing our ability to model complex physical systems. However, it is still challenging to apply computer simulations to physical systems involving multiple and drastically different spatial and temporal scales, such as a concrete example of the thin-film lubrication system. On the one hand, although atomic simulation is used extensively in the investigation of nanoscale phenomena, its size limit is far short to reach macroscale of the system because of the limitation in computer capacity. On the other hand, the conventional continuum mechanics based on finite element method cannot be applied to the nanoscale interfacial region of the system because of the difference between motions of the discrete atoms and collective behaviors of many atoms. The scientific challenge is how to meld an atomistic description that is necessary for treating the highly heterogeneous interfacial region with a continuum description that is adequate for the bulk region of the system.

During the past three decades, a series of effective multi-scale approaches that combine the atomistic and continuum descriptions of fluids, solids and macromoleculars have been developed[9, 10, 11, 12, 13, 14, 15, 16, 17, 18, 19, 20, 21, 22, 23, 24, 25, 26, 27, 28, 29, 30, 31, 32, 33]. A prominent quasistatic multi-scale approach is the quasicontinuum method, in which the whole lattice of materials are coarsely divided by overlaying it with a finite-element mesh to treat the quasistatic evolution of defects at zero temperature[11]. It eliminates the myriad original atoms in favor of many fewer nodes of the mesh. With the correction of free energy on the non-nodal atoms of the mesh, an extension of the quasicontinuum method to treat solid systems at nonzero temperature was developed[18]. The extended quasicontinuum technique or the free-energy corrected hybrid atomistic and coarse-grained method was then successfully applied to treat the two-dimensional thin-film lubrication[34, 35]. Moreover, a hybrid atomistic/continuum modeling for the two-dimensional contact and dry friction between solid substrates was proposed[36, 37]. Recently, for the three-dimensional contact, a multi-scale coupling of the molecular dynamics and the finite element method was reported[38].

In this paper, we develop a multiscale method to treat three-dimensional thin-film lubrication and account for elastic effects of substrates. Section 2 describes the mono-layer molecularly thin-film lubrication system and the multiscale method. Section 3 provides the computational procedure and parameters of the system. The numerical results of the thin-film lubrication system obtained by using the fully atomistic, the several approximate and the multiscale methods are analyzed in section 4. Meanwhile, the validity of the multiscale method in a range of temperature and film-substrate coupling strength and its application to treat the multi-layer molecularly thin-film lubrication systems are further presented. Finally, in Sect. 5, the conclusions are given.

## 2 Three-Dimensional Thin-Film Lubrication System

### 2.1 Fully Atomistic Description

Figure 1(a) is a schematic diagram of the simplified three-dimensional contact, which consists of a face-centered cubic (FCC) crystalline substrate and a mono-layer molecularly thin-film. The upper substrate is assumed as elastic, while the lower substrate is assumed as rigid and reduced to the bottom wall due to the consideration of timesaving computation. The top and bottom walls are assumed to be rigid and to keep crystallographically aligned within the FCC (001) plane. The bottom wall remains stationary in the "laboratory" reference frame; the top wall can be shifted in the $x$- and $y$-directions. The top wall remains parallel with the bottom wall and serves as a handle by which the substrate can be manipulated. Its sliding distance in the $x$-direction is specified by the registry $\alpha$, which is defined as

$$x_i^t = x_i^b + \alpha a, \tag{1}$$

where $x_i^t$ and $x_i^b$ denote the lateral positions of atoms in the top and bottom walls, respectively. $\alpha$ is the distance in the unit of the lattice constant by which the top wall is displaced laterally with respect to the bottom wall. The total configurational energy of the system is then given by

$$\begin{aligned}U(\mathbf{r}^{N_f},\mathbf{r}^{N_s}) = &\tfrac{1}{2}\sum_{i=1}^{N_f}\sum_{j\neq i}^{N_f} u_{ff}(r_{ij}) + \sum_{i=1}^{N_f}\sum_{j=1}^{N_s} u_{fs}(r_{ij}) \\ &+ \tfrac{1}{2}\sum_{i=1}^{N_s}\sum_{j=1}^{N_s} u_{ss}(r_{ij}) + \sum_{i=1}^{N_f}\sum_{j=1}^{N_w} u_{fs}(r_{ij}) \\ &+ \sum_{i=1}^{N_s}\sum_{j=1}^{N_w} u_{ss}(r_{ij}) + \sum_{i=1}^{N_w^t}\sum_{j=1}^{N_w^b} u_{ss}(r_{ij}),\end{aligned} \qquad (2)$$

where $\mathbf{r}^{N_f}$ and $\mathbf{r}^{N_s}$ stand for the collections of $3N_f$ fluid-atom coordinates in the film and $3N_s$ solid-atom coordinates in the substrate. Note that $U$ depends on the $3N_w = 3(N_w^t + N_w^b)$ coordinates of the wall atoms through the last three terms in Eq. (2). The pair interactions are taken to be shifted Lennard-Jones (12,6) potentials

$$u_{ab}(r) = \begin{cases} \phi_{ab}(r) - \phi_{ab}(r_c), & \text{if } r < r_c; \\ 0, & \text{if } r \geq r_c, \end{cases} \qquad (3)$$

where

$$\phi_{ab}(r) = 4\epsilon_{ab}[(\sigma/r)^{12} - (\sigma/r)^6], \quad ab = ff, fs, ss. \qquad (4)$$

The effective diameter $\sigma$ and range $r_c$ are the same for all pairs. Only the depth $\epsilon_{ab}$ of the attractive well depends on the composition of the pair.

## 2.2 Multiscale Description

Figure 1(b) displays a schematic of the hybrid atomistic and coarse-grained treatment of the mono-layer molecular thin-film lubrication. The substrate connected with the top wall is divided into near and far regions, which are depicted by using the atomistic and coarse-grained descriptions, respectively. The coarse-grained far region is covered with a finite-element mesh and connected to the near region of the substrate. The finite-element mesh in Fig.

1(b) consists of "local" tetrahedral elements, whose nodes coincide with a subset of (active) solid substrate and wall atoms[11, 17]. For the local element, a cut-off sphere placed at its "centroid" atom (i.e., the atom nearest the center of inscribed sphere of the element) is embedded itself. In a tetrahedral element, the center of inscribed sphere is written as

$$r_{c\alpha} = \sum_{k=1}^{4} R_{k\alpha} A_k / \sum_{k=1}^{4} A_k, \qquad (5)$$

where $k$ is the index of the nodes, $R_{k\alpha}$ is the nodal coordinates of the element and $A_k$ is the area of the surface of the element opposite the node $k$. As shown in Fig. 1(b), the coarse-grained far region of the substrate is divided into five tetrahedral elements, which include of four congruent and one bigger (see Fig. 2). When an element $e$ is distorted by displacing its nodes, the lattice underlying the element $e$ is assumed to have a homogeneous deformation. In a tetrahedral element, the coordinates $r_{i\alpha}$ of the underlying atom $i$ can be written in terms of the nodal coordinates $(R_{k\alpha})$ as

$$r_{i\alpha} = \sum_{k=1}^{4} V_i(k) R_{k\alpha} / V_e; \alpha = x, y, z, \qquad (6)$$

where $V_e$ denotes the volume of the element $e$ and $V_i(k)$ denotes the volume of the inscribed tetrahedron which has one surface coinciding with the surface of the element $e$ opposite node $k$ and the vertex coinciding with lattice site at $(r_{ix}, r_{iy}, r_{iz})$.

Due to the coarse-graining description the original substrate atoms in the far region is partitioned into two subsets: $N_n$ nodal atoms and $N_q$ non-nodal atoms. Under the assumption of the homogeneous deformation of the element, the contribution of the non-nodal atoms to the free-energy can be transformed into the effective configuration energy of the coarse-grained system. By integrating the Boltzmann factor over the $3N_q$ degrees of freedom of the non-nodal atoms, the effective configurational energy governing the motion of the nodal atoms can be written as[18]

$$V_{eff}(\mathbf{R}^{N_n}) = \sum_{e=1}^{N_e}(N_a^e \tilde{u}_e + N_q^e f_e), \qquad (7)$$

where $\mathbf{R}^{N_n}$ stands for the nodal configuration, $N_e$ for the number of elements, $N_a^e(N_q^e)$ the number of atoms (non-nodal atoms) underlying element $e$. The configurational energy per atom $\tilde{u}_e$ is expressed as

$$\tilde{u}_e = \frac{1}{2}\sum_{j \neq i} u_{ss}(r_{ij}), \qquad (8)$$

where $i$ denotes the centroid atom and $j$ labels atoms that lie within the sphere of radius $r_c$ that is centered on $i$. The Helmholtz energy per non-nodal atom $f_e$ is estimated by using the local harmonic approximation[39] and written as

$$f_e = 3k_B T \ln[\hbar |D|^{1/6}/k_B T], \qquad (9)$$

where the elements of the $3 \times 3$ dynamical matrix are given by $(D)_{kl} = m^{-1}(\partial^2 \tilde{u}_e/\partial x_k \partial x_l)_0 (k, l = 1, 2, 3)$, $k$ and $l$ label Cartesian components ($x_1 = x, x_2 = y, x_3 = z$) of the position of the reference atom, and the subscript 0 signifies that the partial derivative is evaluated at the equilibrium configuration.

The total configurational energy of the hybrid system is then given by

$$\begin{aligned}U_c(\mathbf{r}^{N_f}, \mathbf{r}^{N_s'}, \mathbf{R}^{N_n}) =\ & \tfrac{1}{2}\sum_{i=1}^{N_f}\sum_{j\neq i}^{N_f} u_{ff}(r_{ij}) + \sum_{i=1}^{N_f}\sum_{j=1}^{N_w^b} u_{fs}(r_{ij}) \\ & + \sum_{i=1}^{N_f}\sum_{j=1}^{N_s'} u_{fs}(r_{ij}) + \tfrac{1}{2}\sum_{i=1}^{N_s'}\sum_{j\neq i}^{N_s'} u_{ss}(r_{ij}) \\ & + \sum_{i=1}^{N_s'}\sum_{j=1}^{N_w^b} u_{ss}(r_{ij}) + \tfrac{1}{2}\sum_{i=1}^{N_s'}\sum_{j=1}^{N_s''} u_{ss}(r_{ij}) \\ & + \tfrac{1}{2}\sum_{i=1}^{N_w^t}\sum_{j=1}^{N_s''} u_{ss}(r_{ij}) + \sum_{e=1}^{N_e} N_a^e[\tfrac{1}{2}\sum_{j\neq i} u_{ss}(r_{ij})] \\ & + \sum_{e=1}^{N_e} N_q^e 3k_B T \ln[\hbar |D|^{1/6}/k_B T],\end{aligned} \qquad (10)$$

where $N'_s$ and $N''_s$ stand for the numbers of atoms in the near and the far regions of the substrate, respectively. The band of near region is sufficiently wide, so that fluid atoms do not interact with underlying atoms in the far region. The coarse-grained contribution in the next to last term in Eq. (10) effectively accounts for one half of the interactions between underlying and near-region (bottom wall) atoms. The other half of their interactions is covered in the sixth and seventh terms.

## 2.3  Statistical Thermodynamic Analysis

To compute the thermomechanical properties we employ the analogue of the isothermal-isobaric ensemble and take the usual Gibbs free energy $G$ as its characteristic function described by

$$G = E - TS - AL_zT_{zz}, \qquad (11)$$

where $E$ stands for the internal energy, $S$ for the entropy, $A$ is the area of the system in the $x$-$y$ plane and $L_z$ is the vertical distance between two walls. The change in $G$ under a reversible transformation of the system is written as[40, 41]

$$dG = -SdT + \mu dN_f + \gamma dA - AL_z dT_{zz} + T_{zx}Ad(\alpha a), \qquad (12)$$

where $\mu$ is the chemical potential of the fluid, $\gamma$ is the interfacial tension and $T_{zx}$ is the shear stress.

By means of a standard statistical-mechanical analysis[42, 43], we can obtain the following relationship of the characteristic function $G$ with the isothermal-isobaric partition function $\Delta$

$$G = -k_B T ln\Delta, \qquad (13)$$

where
$$\Delta(T, N_f, T_{zz}, \alpha a) = \sum_{L_z} \exp[(AL_z)T_{zz}/k_B T]Q_{N_f}. \tag{14}$$

In Eq. (14), the canonical partition function $Q_{N_f}$ is expressible in the classical limit as

$$Q_{N_f} = \frac{1}{h^{3(N_f+N_s)}} \frac{1}{N_f!} \int d\mathbf{p}^{N_f} \int d\mathbf{r}^{N_f} \int d\mathbf{p}^{N_s} \int d\mathbf{r}^{N_s} \exp(-H/k_B T), \tag{15}$$

where $h$ is the Planck's constant. The Hamiltonian $H$ is written as

$$H = \sum_{i=1}^{N_f} \mathbf{p}_i^2/2m_f + \sum_{i=1}^{N_s} \mathbf{p}_i^2/2m_s + U(\mathbf{r}^{N_f}, \mathbf{r}^{N_s}), \tag{16}$$

where $\mathbf{p}^{N_f}$ and $\mathbf{p}^{N_s}$ represent the momenta conjugate to the coordinates $\mathbf{r}^{N_f}$ and $\mathbf{r}^{N_s}$, respectively. After the integral on momenta, the Eq. (15) can be rewritten as

$$Q_{N_f} = \frac{1}{N_f!} \Lambda_f^{3N_f} \Lambda_s^{3N_s} Z_{N_f}, \tag{17}$$

where $\Lambda_f = \sqrt{2\pi m_f T}/h$, $\Lambda_s = \sqrt{2\pi m_s T}/h$ and the configuration integral $Z_{N_f}$ is

$$Z_{N_f} = \int d\mathbf{r}^{N_f} \int d\mathbf{r}^{N_s} \exp(-U/k_B T). \tag{18}$$

From Eqs. (12)-(16), we obtain for the shear stress

$$AT_{zx} = \frac{\partial G}{\partial \alpha a} = -\frac{k_B T}{\Delta} \frac{\partial \Delta}{\partial \alpha a} = -k_B T \frac{\sum_{L_z} \exp(AL_z T_{zz}/k_B T) \frac{\partial Z_{N_f}}{\partial \alpha a}}{\sum_{L_z} \exp(AL_z T_{zz}/k_B T) Z_{N_f}}, \tag{19}$$

Using the transformation of variable

$$x'_i = x_i - \alpha a z_i/L_z \tag{20}$$

we can rewrite the configuration integral $Z_{N_f}$ as

$$Z_{N_f} = \Pi_{i=1}^{N_f+N_s} \int d\mathbf{r}^{N_f} \int d\mathbf{r}^{N_s} \exp(-U[x'_i + \alpha a z_i/L_z, y_i, z_i]/k_B T). \tag{21}$$

Its differentiation with respect to $(\alpha a)$ is written as

$$\frac{\partial Z_{N_f}}{\partial \alpha a} = -\frac{1}{k_B T L_z} \int d\mathbf{r}^{N_f} \int d\mathbf{r}^{N_s} \exp(-U/k_B T) \frac{\partial U}{\partial \alpha a}. \tag{22}$$

Finally, we obtain the shear stress $T_{zx}$ of the fully atomistic system

$$
\begin{aligned}
T_{zx} = & \tfrac{1}{2A} \sum_{i=1}^{N_f} \sum_{j \neq i}^{N_f} < u'_{ff}(r_{ij}) x_{ij} z_{ij}/(r_{ij} L_z) > \\
& + \tfrac{1}{A} \sum_{i=1}^{N_f} \sum_{j=1}^{N_s} < u'_{fs}(r_{ij}) x_{ij} z_{ij}/(r_{ij} L_z) > \\
& + \tfrac{1}{2A} \sum_{i=1}^{N_s} \sum_{j=1}^{N_s} < u'_{ss}(r_{ij}) x_{ij} z_{ij}/(r_{ij} L_z) > \\
& + \tfrac{1}{A} \sum_{i=1}^{N_f} \sum_{j=1}^{N_w} < u'_{fs}(r_{ij}) x_{ij} z_{ij}/(r_{ij} L_z) > \\
& + \tfrac{1}{A} \sum_{i=1}^{N_s} \sum_{j=1}^{N_w} < u'_{ss}(r_{ij}) x_{ij} z_{ij}/(r_{ij} L_z) > \\
& + \tfrac{1}{A} \sum_{i=1}^{N_w^t} \sum_{j=1}^{N_w^b} < u'_{ss}(r_{ij}) x_{ij} z_{ij}/(r_{ij} L_z) >,
\end{aligned}
\quad (23)
$$

where $u'_{ab} = \frac{\partial u_{ab}}{\partial r_{ij}} = -24\epsilon_{ab}(2\sigma^{12}/r_{ij}^{13} - \sigma^6/r_{ij}^7)$. The angular brackets $< W >$ signify the ensemble average

$$< W > = \sum_{L_z} \int d\mathbf{r}^{N_f} \int d\mathbf{r}^{N_s} P(\mathbf{r}^{N_f}, \mathbf{r}^{N_s}; L_z) W(\mathbf{r}^{N_f}, \mathbf{r}^{N_s}; L_z), \quad (24)$$

where $P$ is the isothermal-isobaric distribution function.

For the hybrid system, by carrying out the integrals over the formal momentum space $(\mathbf{p}^{N_f}, \mathbf{p}^{N'_s}, \mathbf{P}^{N_n})$, we obtain the canonical partition function $Q_{N_f,c}$ in the classical limit as

$$Q_{N_f,c} = \frac{1}{h^{3(N_f+N'_s+N_n)}} \frac{1}{N_f!} \int d\mathbf{p}^{N_f} \int d\mathbf{r}^{N_f} \int d\mathbf{p}^{N'_s} \int d\mathbf{r}^{N'_s} \int d\mathbf{P}^{N_n} \int d\mathbf{R}^{N_n} \exp(-\tilde{H}_c/k_B T), \quad (25)$$

where

$$
\begin{aligned}
\tilde{H}_c = & \sum_{i=1}^{N_f} \mathbf{p}_i^2/2m_f + \sum_{i=1}^{N'_s} \mathbf{p}_i^2/2m_s + \tfrac{1}{2}\mathbf{P}^T \mathbf{M}^{-1} \mathbf{P} \\
& + U_c(\mathbf{r}^{N_f}, \mathbf{r}^{N'_s}, \mathbf{R}^{N_n}).
\end{aligned}
\quad (26)
$$

where $\mathbf{P}$ is the $3N_n$-dimensional column vector of momenta conjugate to the nodal coordinates $\mathbf{R}^{N_n}$ and $\mathbf{M}^{-1}$ is the effective mass matrix, which depends on the transformation in Eq. (6)[17]. After the integral on momenta, the Eq. (25) can be rewritten as

$$Q_{N_f,c} = \frac{1}{h^{3N_n}} \frac{1}{N_f!} \Lambda_f^{3N_f} \Lambda_s^{3N'_s} \int d\mathbf{P}^{N_n} \exp(-\mathbf{P}^T \mathbf{M}^{-1} \mathbf{P}/2k_B T) Z_{N_f,c}, \quad (27)$$

where

$$Z_{N_f,c} = \int d\mathbf{r}^{N_f} \int d\mathbf{r}^{N'_s} \int d\mathbf{R}^{N_n} \exp(-U_c/k_B T). \tag{28}$$

In a similar derivation, we obtain the shear stress $T_{zx,c}$ of the hybrid system

$$\begin{aligned}
T_{zx,c} =\ & \tfrac{1}{2A} \sum_{i=1}^{N_f} \sum_{j\neq i}^{N_f} <u'_{ff}(r_{ij})x_{ij}z_{ij}/(r_{ij}L_z)> \\
& + \tfrac{1}{A} \sum_{i=1}^{N_f} \sum_{j=1}^{N_w^b} <u'_{fs}(r_{ij})x_{ij}z_{ij}/(r_{ij}L_z)> \\
& + \tfrac{1}{A} \sum_{i=1}^{N_f} \sum_{j=1}^{N'_s} <u'_{fs}(r_{ij})x_{ij}z_{ij}/(r_{ij}L_z)> \\
& + \tfrac{1}{2A} \sum_{i=1}^{N'_s} \sum_{j\neq i}^{N'_s} <u'_{ss}(r_{ij})x_{ij}z_{ij}/(r_{ij}L_z)> \\
& + \tfrac{1}{2A} \sum_{i=1}^{N'_s} \sum_{j=1}^{N''_s} <u'_{ss}(r_{ij})x_{ij}z_{ij}/(r_{ij}L_z)> \\
& + \tfrac{1}{A} \sum_{i=1}^{N'_s} \sum_{j=1}^{N_w^b} <u'_{ss}(r_{ij})x_{ij}z_{ij}/(r_{ij}L_z)> \\
& + \tfrac{1}{2A} \sum_{i=1}^{N_w^t} \sum_{j=1}^{N''_s} <u'_{ss}(r_{ij})x_{ij}z_{ij}/(r_{ij}L_z)> \\
& + \tfrac{1}{A} \sum_{e=1}^{N_e} N_a^e [\tfrac{1}{2} \sum_{j\neq i} <u'_{ss}(r_{ij})x_{ij}z_{ij}/(r_{ij}L_z)>] \\
& + \tfrac{1}{A} \sum_{e=1}^{N_e} N_q^e [\tfrac{k_B T}{2} <\tfrac{1}{|D|}\tfrac{\partial |D|}{\partial \alpha a}>].
\end{aligned} \tag{29}$$

# 3  Computational Procedure

We survey the behaviors of the system under reversible shearing at a given atomistic number $N_f + N_s$ or $N_f + N'_s + N_n$, temperature $T$ and normal stress $-T_{zz}$, and investigate effects of elasticity of the substrate, temperature and film-substrate coupling strength on the shear stress. To compute the thermomechanical properties we perform isothermal-isobaric Monte Carlo (MC) simulations[44]. Shearing process, which is regarded as a quasistatic (reversible), is performed by increasing $\alpha$ gradually in small increments $\Delta\alpha$. For the register $\alpha = 0$, the initial film atoms are randomly placed between two solid surfaces. The initial configuration of the system for the register $\alpha \neq 0$ is taken from the last configuration of the previous register $\alpha - \Delta\alpha$ combined with a linear increment ($0 \leq \Delta x(z) \leq \Delta\alpha$) of the several rows nearest the top wall. One MC cycle consists of diffusive steps of atoms and nodes with $L_z$ fixed or exchanged. To facilitate the equilibration of the fluid film, the fluid atoms are subjected to a cycle of the normal move and several cycles of the large move. We generate a total of $M$ MC cycles, discard the first half of these, and evaluate ensemble averages over the remaining half. At some transition points from positive shear stresses to the negative ones or vice versa, we extend the total MC cycles to $2M$ for reaching the quasistatic states of the system.

For the simulations, Table I lists the values of the various parameters for the fully atomistic and multiscale simulations. Numerical values are expressed in dimensionless units based on the Lennard-Jones parameters for the solid-solid interaction: distance is expressed in units of $\sigma$; energy in units of $\epsilon_{ss}$; stress in units of $\epsilon_{ss}/\sigma$; temperature in units of $\epsilon_{ss}/k_B$.

## 4 Results

### 4.1 Fully Atomistic Treatment of the Substrate

We first investigate tribological behaviors of the fully atomistic version of the thin-film lubrication system as shown in Fig. 1(a). Figure 3 displays the shear stress $T_{zx}$ and the mean separation $L_z$ as a function of the register $\alpha$ at $\epsilon_{fs} = 1/9$, $T = 0.15$ and $T_{zz} = -1.0$. It is noted that both forward and backward processes in $\alpha$ are periodic with a length 1. The forward process composites two elastic(stick) zones ($0 \leq \alpha \leq 0.6$, $0.7 \leq \alpha \leq 1$) and a transition(slip) zone ($0.6 \leq \alpha \leq 0.7$). At $\alpha = 0$, fluid atoms converge to a stable FCC crystal structure with solid atoms of the substrate and the bottom wall. The perfect symmetric atomistic structure of the system in the equilibrium state leads to $T_{zx} = 0$. As $\alpha$ increases gradually in the range $0 < \alpha \leq 0.6$, the substrate connected with the bottom wall through the film just likes a whole crystal to generate an elastic shear behavior in the $x-$ direction. Even when $\alpha = 0.5$, the substrate presenting the elastic shear behavior postpones the appearance of the transition process in the system. As shown in Fig. 4(a), although solid atoms in two surfaces and fluid atoms in the thin film are still matched to the FCC crystal structure, the solid atoms in the substrate have the right inclination to those in the bottom wall. The fluid atoms in the film are thus confined in the interstices. A fluid atom in its interstice is closer to the upper solid atom which lies to its left and the lower one which lies to its right. A reversed shear stress (in the negative $x-$ direction) on the upper surface is generated by the thin fluid film, transmitted to the top wall and counterbalanced by the external stress holding the top wall in the shifting place. Meanwhile, as the shear stress in the film increases, the interstices are compressed slender, and then the effective film thickness increases along with the mean separation $L_z$. As $\alpha$ varies in the range $0.6 \leq \alpha \leq 0.7$, the shear stress and the effective film thickness reach the largest values. The atoms in the film or in the upper surface of the substrate abruptly step over the atoms touching with them. To complete the relative striding of the atoms in the system, the substrate must have a sharp climbing to provide the vertical space near the film. Since the slip zone is very small, the atoms in the thin film are almost regular and cannot rejoin the fluidic state. When $\alpha$ increases further in the range $0.7 \leq \alpha \leq 1$, the systematic shear stress curve has a central rotational symmetry with that at $1 - \alpha$. An elastic shear behavior is generated in the substrate connected with the bottom wall through the film. As shown in Fig. 4(b), the solid atoms in two surfaces and fluid atoms in the thin film are matched to a new FCC crystal structure. The solid atoms in the substrate have the left inclination to those in the bottom wall, so the fluid atoms in the film are confined in the interstices. A fluid atom in its interstice is closer to the upper solid atom which lies to its right and the lower one which lies to its left. A shear stress (in the $x-$ direction) on the upper surface is generated by the thin fluid film, transmitted to the top wall and counterbalanced by the external stress holding the top wall in the shifting place. Meanwhile, as the shear stress in the film decreases, the interstices are changed stubbier, and then the effective film thickness decreases along with the mean separation $L_z$. When the register finally reaches $\alpha = 1$, the system returns to the same state as it is at $\alpha = 0$. The solid atoms in the substrate have a relative sliding distance $a$ to those in the bottom wall. As $\alpha$ increases from 1 to 1.2, the above profiles in the range $(0 < \alpha \leq 0.2)$ are repeated. In the reversed slip process, shearing from $\alpha = 1.2$ to $\alpha = 0$ presents the same stick-slip behavior as shearing in the forward direction. The shear stress profile is central rotational symmetric with that in the forward process. In the zone away from the transition process, shearing carries the system in reverse through precisely the same states as shearing in the forward direction. However, due to the elastic effects of the substrate, the states near the slip zone in the backward process do not coincide with those in the forward process, which forms a hysteresis loop.

## 4.2 Elastic Effects of the Substrate and Its Approximate Treatment

To determine the importance of a proper accounting for elasticity of the substrate in the thin-film lubrication system, we take 4 modeling tribological systems containing the successively enlarged elastic region of the substrate with 0-layers near region [wholly rigid substrate], 8-layers near region [partly rigid substrate], 16-layers near region [partly rigid substrate] and 22-layers near regions [fully atomistic substrate]. In the systems, except for the fixed $N_f$, $N_w^t$ and $N_w^b$ as given in Table I, $N_s$ is changed as $162l$, where $l = 0, 8, 16, 22$. The shear stress profiles of the systems at $\epsilon_{fs} = 1/9$, $T = 0.15$ and $T_{zz} = -1.0$ are shown in Fig. 5.

For the tribological system with wholly rigid substrate, the shear stress of the system cannot be relaxed in the rigid substrate and is directly acted on the top wall. The thermal elastic coefficient reaches a large value ($c_0 = \Delta T_{zx}/\Delta \alpha = 8.8$), which reveals that the system loses elastic effects of the substrate. The process of slipping over the atoms in the film or in the bottom wall appears early. The stick zone ($0 \leq \alpha \leq 0.2$, $0.8 \leq \alpha \leq 1$) in the forward process is narrow, but the slip zone ($0.2 \leq \alpha \leq 0.8$) is wide. In the stick zone, as shown in Fig. 6(a)/(c), the solid atoms in the top wall have the right/left inclination to those in the bottom wall. The fluid atoms in the film are confined in the interstices. The solid atoms in two surfaces and the fluid atoms in the film are matched to the FCC crystal structure. In the slip zone, as shown in Fig. 6(b), the film atoms between two surfaces are irregular and rejoin the fluidic state. This transition state does not appear in the fully atomistic version of the tribological system. Meanwhile, shearing from $\alpha = 1.2$ to $\alpha = 0$ carries the system in reverse through precisely the same states as shearing in the forward direction. The hysteresis loop in the fully atomistic version of the tribological system vanishes in the system due to the inexistence of the elastic substrate.

For the tribological system with partly rigid substrates, the thermal elastic coefficient ($c_l = \Delta T_{zx}/\Delta \alpha$) decreases gradually with the increasing of the thickness $l$ of the elastic region of the substrate. The largest shear stress $T_{zx}^*$ has a slight increasing in comparing with that for the system with $l = 0$ and approximately keeps a constant when $l \geq 8$. The stick zones are enlarged and the thermal elastic constants $c_l$ decreases. The slip zones are narrowed, so that the transition process from the positive $T_{zx}^*$ to the negative one becomes more and more quickly. However, the shear-stress profile and atomistic structures depend on the thickness $l$ of the elastic region of the substrate. For the tribological system with 8-layers near region, the shear-stress profile and atomistic structures possess similar behaviors with those for the tribological system with wholly rigid substrate. The stick-slip behavior in the forward process is performed reversibly and coincides with that in the backward process. For the tribological system with 16-layers near region, the shear-stress profile and atomistic structures are similar to those for the fully atomistic version of the tribological system. The states away from the slip zone in the forward process coincide with those in the backward process, but the states near the slip zone do not coincide with those in the forward process. It leads to a hysteresis loop in the forward and backward processes.

Therefore, the approximate treatment [wholly ($l = 0$) or partly rigid ($l = 8$)] of the substrate in the thin-film lubrication system can only predict the approximate and partial tribological features in the fully atomistic model. To generate the qualitative tribological behavior with a hysteresis loop in the fully atomistic model, the elastic substrate of the thin-film lubrication system must have the adequate thickness $l \geq 16$. In comparison with the tribological system with the wholly rigid substrate, the hysteresis loop is almost a signature of the elastic response of substrate in fully atomistic version of the thin-film lubrication system.

## 4.3 Hybrid Treatment of the Substrate

As shown in Fig. 1(b), the substrate of the tribological system is treated as the hybrid description with the local elements. The parameters for the multiscale simulation are given in Table I. By using the multiscale method, the shear stress $T_{zx,c}$ and the mean separation $L_{z,c}$ of the system at $\epsilon_{fs} = 1/9$, $T = 0.15$ and $T_{zz} = -1.0$ are determined and compared with those obtained from the fully atomistic simulation in Fig. 7. Both $T_{zx,c}$ and $L_{z,c}$ equal to the "exact" (fully atomistic) values in the forward and backward processes. The stick-slip phenomenon and the corresponding physical mechanism based on the atomistic structures are also coincident with those in the fully atomistic version of the tribological system. It reveals that the elastic response of substrate can be effectively accounted for the hybrid scheme. Meanwhile, the computing time of the multiscale simulation based on the hybrid scheme is only about 1/5 of that of the fully atomistic simulation. So, in comparison with the fully atomistic simulation, the multiscale simulation based on the hybrid scheme can greatly save computing time and increase the computational efficiency.

To further confirm the validity of the multiscale simulation of the thin-film lubrication system, the temperature $T$ and the film-substrate coupling strength $\epsilon_{fs}$ are chosen as the variable parameters of the system in following subsections.

### 4.3.1 Thermal Effects

Figure 8 displays the shear stress profiles of the system at several temperatures ($T = 0.05$, $0.15$, $0.25$) and fixed $\epsilon_{fs} = 1/9$ and $T_{zz} = -1.0$. $T_{zx,c}$ generated by the multiscale simulation is also in excellently agreement with $T_{zx}$ by the fully atomistic simulation. The thermal elastic coefficient $c$ decreases as the temperature increases. It reveals that the stiffness of the tribological system is reduced, so that the largest shear stress $T_{zx}^*$ decreases and the transition process appears earlier. Meanwhile, the hysteresis loops in the forward and backward processes are smaller due to the losing/increasing of the stiffness/elasticity of the substrate.

### 4.3.2  Effects of the film-substrate coupling strength

Figure 9 displays the shear stress profiles of the system at several film-substrate coupling strengths $\epsilon_{sf}$=1/3, 1/9 and 1/12 and fixed $T = 0.15$ and $T_{zz} = -1.0$. $T_{zx,c}$ generated by the multiscale simulation is also in excellently agreement with $T_{zx}$ by the fully atomistic simulation. The thermal elastic coefficients $c$ are almost constant in the process of changing the film-substrate coupling strength $\epsilon_{sf}$. It reveals that in the tribological systems the intermolecular interaction between the substrate and the bottom wall is dominant. The film-substrate coupling strength, which describes the intermolecular interaction between the thin film and the substrate or the bottom wall, hardly affects the stiffness of the tribological system. For the film-substrate coupling strength $\epsilon_{sf} = 1/3$, $\epsilon_{sf}$ is equal to $\epsilon_{ss}$, i.e., the intermolecular interaction between the film and the substrate is the same with that of the substrate atom. It directly leads to the increasing of the adhesive force of the fluid atoms to the solid surfaces and the delaying of the slip process. A large hysteresis loop appears in the forward and backward processes. For the film-substrate coupling strength $\epsilon_{sf}$=1/12, $\epsilon_{sf}$ is smaller than $\epsilon_{ss}$. The tribological system has the weaker adhesive force than that for $\epsilon_{sf}$=1/9. It leads to that both the systems with $\epsilon_{sf}$=1/9 and 1/12 have the similar tribological behaviors.

From the above results, we can conclude that the multiscale method is effective to predict the tribological behaviors of the thin-film lubrication systems in the range of parameters (temperature and film-substrate coupling strength).

## 4.4 Application of the multiscale method

In the following, as an application of the multiscale method, we take three cases of three-dimensional tribological systems with the mono-layer, the two-layer and the three-layer molecularly thin-films to investigate tribological behaviors. Schematics of the hybrid atomistic and coarse-grained treatment of the two-layer and the three-layer molecularly thin-film lubrication systems are shown in Fig. 10. Using the multiscale method, we focus on the effects of loads on the tribological behaviors of the systems under reversible shearing at a given atomistic number $n_l N_f + N_s' + N_n$, temperature $T = 0.15$ and the load $-T_{zz}$, where $n_l$=1, 2 and 3 correspond to the mono-layer, the two-layer and the three-layer films, respectively. The film-substrate coupling strength is fixed as $\epsilon_{sf} = 1/9$.

Figure 11(a) presents the shear stress $T_{xz}$ versus $\alpha$ curves under several loads $-T_{zz}$ for the mono-layer molecularly thin-film. The stick-slip processes with the hysteresis loops appear when the load $-T_{zz} \geq 0.5$. At the lightest load $-T_{zz} = 0.1$, the curve in the forward process exactly coincides with that in the backward process. The molecular tribological mechanism based on the stick-slip processes was previously analyzed in Sect. 4.1 and Fig. 4, where the load $-T_{zz}$=1.0 is taken as an example. The thermal elastic coefficient $c_{Tzz}$ increases monotonously as the load $-T_{zz}$ increases, which leads to the larger shear stress and hysteresis loop in the forward and backward processes. The heavier the load, the more compressed the confined film. It generates stronger intermolecular forces to prevent sliding between two solid surfaces.

Figures 11(b) presents the shear stress $T_{xz}$ versus $\alpha$ curves under several loads $-T_{zz}$ for the two-layer molecularly thin-film. In comparison with the mono-layer film tribological model, on the one hand, the equilibrium state with a stable FCC crystal structure between the thin-film and the two sur-faces is changed from $\alpha = 0/1$ to $-0.5/0.5$, and consequently the stick-slip period in the $T_{xz}-\alpha$ profile is changed from $[0, 1]$ to $[-0.5, 0.5]$. The stick-slip process with a hysteresis loop appears when the load $-T_{zz} \geq 2.0$. At the lighter loads $-T_{zz} = 0.5$ and $1.0$, the curves in the forward processes exactly coincide with those in the backward processes. As shown in Fig. 12(a)(c), since the fluid atoms (open red spheres) in the second-layer film serving as the solid atoms to make the stable FCC crystal structures, the solid atoms in the substrate at $\alpha = -0.2$ and $0.2$ have the right and the left inclination to those in the bottom wall, respectively. The fluid atoms are confined in the interstices. A transitional state appears at $\alpha = 0$ and postpones the slipping process. The fluid atoms (filled and open red spheres) in the two-layer thin-film are placed irregularly between the two surfaces as shown in Fig. 12(b). On the other hand, the thermal elastic coefficient $c_{Tzz}$ increases monotonously with the load $-T_{zz}$ for the same reason that the shear stress and the hysteresis loop increases with the load. The difference is that $c_{Tzz}$ and $T_{xz}$ at a fixed $-T_{zz}$ is markedly reduced from the mono-layer film system to the two-layer film system. The thicker the confined film, the weaker the intermolecular force between two solid surfaces. It leads to promote the sliding between two surfaces.

Figures 11(c) presents the shear stress $T_{xz}$ versus $\alpha$ curves under several loads $-T_{zz}$ for the three-layer molecularly thin-film. In comparison with the two-layer film tribological model, the equilibrium state and the stick-slip period return back to those of the mono-layer film system. As shown in Fig. 13(a)(c), the second-layer fluid atoms (open red spheres) and the third-layer fluid atoms (open read squares) in the thin-film at $\alpha = 0.3$ and $0.7$ serve as the solid atoms to form the stable FCC crystal structures. The fluid atoms (filled and open red spheres, and open squares) in the three-layer thin-film at $\alpha = 0.5$ are placed irregularly between the two surfaces as shown in Fig. 13(b). Most tribological behaviors are similar to those of two-layer film system. The difference is that the thermal elastic coefficient $c_{Tzz}$ and the shear stress $T_{xz}$ at a fixed load is slightly reduced from the two-layer film system to the three-layer film system. It reveals that the three-layer film replacing the two-layer film can only slightly reduce the intermolecular force between two solid surfaces and promote the sliding between them.

Dependence of the maximal friction force $T_{xz}^* A$ (the product of the maximal shear stress $T_{xz}^*$ and the surface area $A$) on a normal force $-T_{zz}A$ represents systematic tribological properties. Figure 14 displays the maximal shear stress $T_{xz}^*$ versus the load $-T_{zz}$ curves for the mono-layer, the two-layer and the three-layer film systems. At a fixed load, $T_{xz}^*$ of the mono-layer film system is remarkably larger than those of the two-layer and the three-layer film systems. $T_{xz}^*$ of the two-layer film system is slightly larger than that of the three-layer film system. For the three cases, $T_{xz}^*$ rises approximately linearly as the load $-T_{zz}$ increases. According to the Amontons-Coulomb friction law, the systematic static friction coefficient is defined as $\mu = -T_{xz}^* A / T_{zz} A = -T_{xz}^*/T_{zz}$. For the mono-layer film system, even the load $-T_{zz}$ approaches to zero, a large intermolecular interaction between two solid surfaces would be overcome to realize the relative sliding of them, i.e., $T_{xz}^* > 0$. So the mono-layer molecularly thin-film lubrication does not satisfy the Amontons-Coulomb friction law as shown in Fig. 14. However, for the two-layer and the three-layer film systems, since the thicker films obstruct the intermolecular interactions between two solid surfaces, they can approximately satisfy the Amontons-Coulomb friction law as shown in Fig. 14. To generate an uniform law to represent the dependence of $T_{xz}^*$ on $-T_{zz}$ for the three cases, an improved systematic static friction coefficient is proposed as $\mu = -\Delta T^*_{xz}A/\Delta T_{zz}A = -\Delta T^*_{xz}/\Delta T_{zz}$. Using this revised definition, we determine $\mu_1$=0.82, $\mu_2$=0.65 and $\mu_3$=0.58 for the mono-layer, the two-layer and the three-layer film systems, respectively. Monotonously decreasing of the systematic static friction coefficient reveals the functions of reducing friction and wear of the multi-layer molecularly thin-film lubrication.

# 5 Conclusions

We investigate effects of the elastic substrate on tribological properties of three-dimensional mono-layer molecularly thin-film lubrication and propose a multiscale method, which combines an atomistic description of the near region with a coarse-grained description of the far region of the solid substrate, to treat the thin-film lubrication. The coarse-graining reduces the original Hamiltonian of the far region to an effective one, which not only governs the nodes of the finite element meshes as pseudo-atoms but also takes account of the harmonic contributions of non-nodal atoms in the finite element meshes. The multiscale method is applied to simulate reversible shearing of the thin-film lubrication and compare with its several approximation versions related to the substrate. The validity of the Amontons-Coulomb friction law on the multi-layer molecularly thin-film lubrication is further tested in the multiscale simulation based on the hybrid scheme. The conclusions of the investigations are summarized as follows.

(1) Neglecting completely the elastic substrate (wholly rigid treatment of the substrate) misses the snapping character in the transition process of the thin-film lubrication with the elastic substrate (fully atomistic version of the system) and the hysteresis loop of the shear-stress profile in the forward and backward processes.

(2) introducing partly elasticity near the fluid-solid interface improves the shape of the shear-stress profile of the thin-film lubrication, which depends on the thickness $l$ (the molecular layer number) of the elastic substrate. The tribological properties generated in the system with the thickness $l \leq 8$ are similar to those with wholly rigid substrate ($l = 0$). When the thickness $l \geq l_c = 16$ ($l_c$ is a critical thickness), the tribological properties of the system are qualitatively consistent with those for fully atomistic version of the thin-film lubrication ($l = 22$).

(3) both the shear-stress and the mean-separation profiles determined by using the multiscale method are in excellently agreement with those obtained from the fully atomistic simulation. The elastic response of substrate can be effectively rendered in the hybrid scheme. The multiscale method can successfully predict the tribological properties of the thin-film lubrication system in a range of parameters (temperature and film-substrate coupling strength), greatly save computing time and increase the computational efficiency.

(4) the systematic static friction coefficient monotonously decreases as as the molecular layer number in the fluid film increases. In comparison with the mono-layer molecularly thin-film lubrication, the multi-layer molecularly thin-film lubrication plays the functions of reducing friction and wear of the system by decreasing the systematic static friction coefficient.

**Acknowledgments** This research is supported by the National Natural Science Foundation of China through the Grants No. 11172310 and No. 11472284 and the CAS Strategic Priority Research Program XDB22040403. The author thanks the National Supercomputing Center in Tianjin for assisting in the computation.

Table I. Parameters of the fully atomistic and multiscale simulations of the thin-film lubrication system.

| |
|---|
| Number of film atoms $N_f = 162$ |
| Number of top wall atoms $N_w^t = 162$ |
| Number of bottom wall atoms $N_w^b = 162$ |
| Number of atoms in a FCC cell $n_c = 4$ |
| Total number of substrate atoms $N_s = 9 \times 9 \times 11 \times n_c = 3564$ |
| Number of near-region atoms $N_s' = 9 \times 9 \times 2 \times n_c = 648$ |
| Number of far-region atoms $N_s'' = 9 \times 9 \times 9 \times n_c = 2916$ |
| Number of local elements $N_e = 5$ |
| Number of nodes for local elements $N_n = 1$ |
| Density of substrate $\rho = 1.1$ |
| Lattice constant of substrate $a = (\frac{4}{\rho})^{1/3} = 1.538$ |
| Area of contact $A = L_x \times L_y = 9a \times 9a = 81a^2$ |
| Substrate-substrate Lennard-Jones depth $\epsilon_{ss} = 1$ |
| Film-film Lennard-Jones depth $\epsilon_{ff} = 1/9$ |
| Cutoff radius $r_c = 2.5$ |
| Total number of Monte Carlo cycles $M = 10^5$ |

FIGURE CAPTIONS

Figure 1. Schematic diagram of three-dimensional mono-layer molecularly thin-film lubrication system with periodic boundary conditions in $x-$ and $y-$ directions. The wall, the substrate and the fluid atoms are represented by green, blue and red spheres, respectively. (a) Fully atomistic description; (b) partial coarse graining of far region of the substrate with local elements. The coarse grained far region of the substrate is represented by gray spheres. All atoms are depicted in their initial configuration at $\alpha = 0$.

Figure 2. Schematic diagram of dividing a cube to constitute coarse-graining mesh in the far region of the substrate. The tetrahedral elements (1),(3),(4) and (5) are congruent and smaller than the tetrahedral element (2).

Figure 3. Shear stress profile ($T_{zx}$ vs register $\alpha$) and the mean separation profile ($L_z$ vs register $\alpha$) of the thin-film lubrication system with $\epsilon_{fs} = 1/9$ at $T$=0.15 and $T_{zz}$=-1.0(forward, by solid line with circles; backward, by dashed line with triangles).

Figure 4. Atomistic structures ($x - z$ side view and $x - y$ bottom view) of the tribological system at (a) $\alpha = 0.6$; (b) 0.7. Labels are same as those of Fig. 1.

Figure 5. Shear-stress profile ($T_{zx}$ versus register $\alpha$) for the models with the elastic substrates of different thickness $l$ at $\epsilon_{fs} = 1/9$, $T$=0.15 and $T_{zz}$=-1.0 (forward, by solid line; backward, by dashed line). Wholly rigid substrate/0-layer near region (diamonds); 8-layer near region (triangles); 16-layer near region (squares); fully atomistic/22-layer near region (circles).

Figure 6. Atomistic structures ($x - z$ side view and $x - y$ bottom view) of the modeling tribological system at (a) $\alpha = 0.2$; (b) 0.5; (c) 0.8. Labels are same as those of Fig. 1.

Figure 7. Shear stress profiles profiles at $\epsilon_{fs} = 1/9$, $T=0.15$ and $T_{zz}=-1.0$ generated in fully atomistic and multiscale simulations. Fully atomistic profile (forward, by solid line; backward, by dashed line); multiscale profile (forward, by filled triangles; backward, by hollow triangles).

Figure 8. Same as figure 7, except for (a) $T=0.05$, (b) 0.15 and (c) 0.25.

Figure 9. Same as figure 7, except for (a) $\epsilon_{fs}=1/3$, (b) 1/9 and (c) 1/12.

Figure 10. Schematic diagram of three-dimensional (a) two-layer and (b) three-layer molecularly thin-film lubrication system with periodic boundary conditions in $x-$ and $y-$ directions. Labels are same as those of Fig. 1(b).

Figure 11. Shear stress $T_{xz}$ versus $\alpha$ curves under several loads $-T_{zz}$ for (a) the mono-layer, (b) the two-layer and (c) the three-layer molecularly thin-film.

Figure 12. Atomistic structures ($x-z$ side view and $x-y$ bottom view) of the two-layer molecularly tribological system at (a) $\alpha = -0.2$; (b) 0.0; (c) 0.2. The fluid atoms in the second layer are represented by open red spheres. Other labels are same as those of Fig. 1.

Figure 13. Atomistic structures ($x-z$ side view and $x-y$ bottom view) of the three-layer molecularly tribological system at (a) $\alpha = 0.3$; (b) 0.5; (c) 0.8. The fluid atoms in the second and the third layers are represented by open red spheres and squares, respectively. Other labels are same as those of Fig. 1.

Figure 14.The maximal shear stress $T_{xz}^*$ versus the load $-T_{zz}$ for the mono-layer, the two-layer and the three-layer fluid films. The systematic static friction coefficient $\mu_l$ of the $l$-layer fluid film is given.

**Figure 1**
Click here to download Figure: Figure1.eps

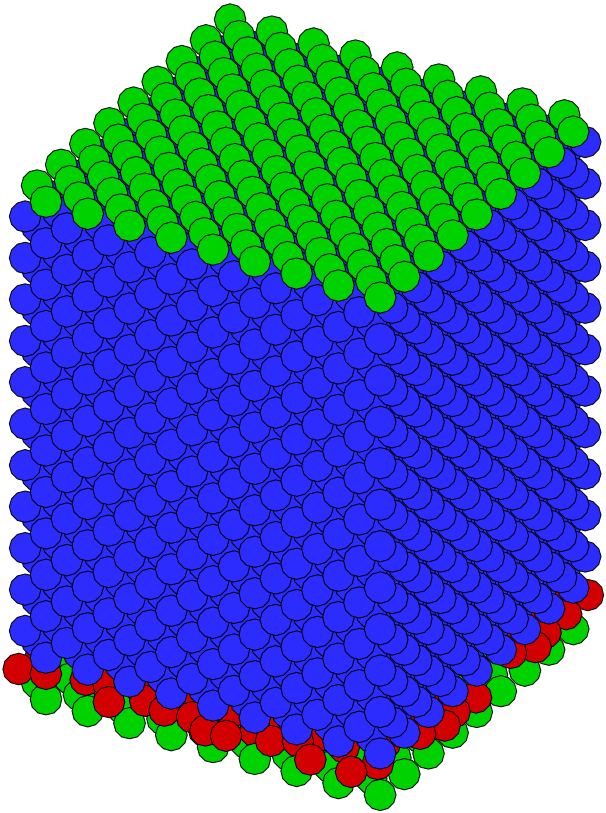 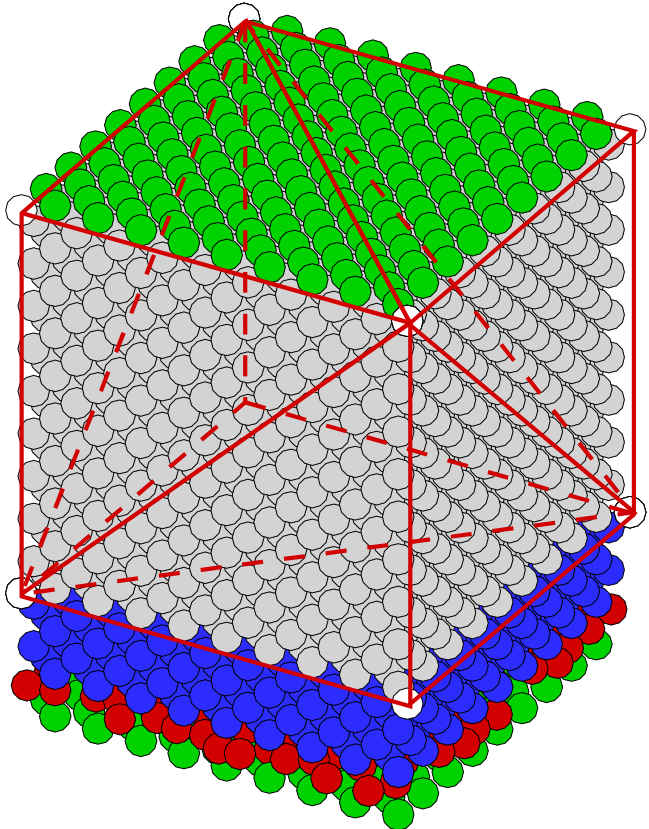

(a) (b)

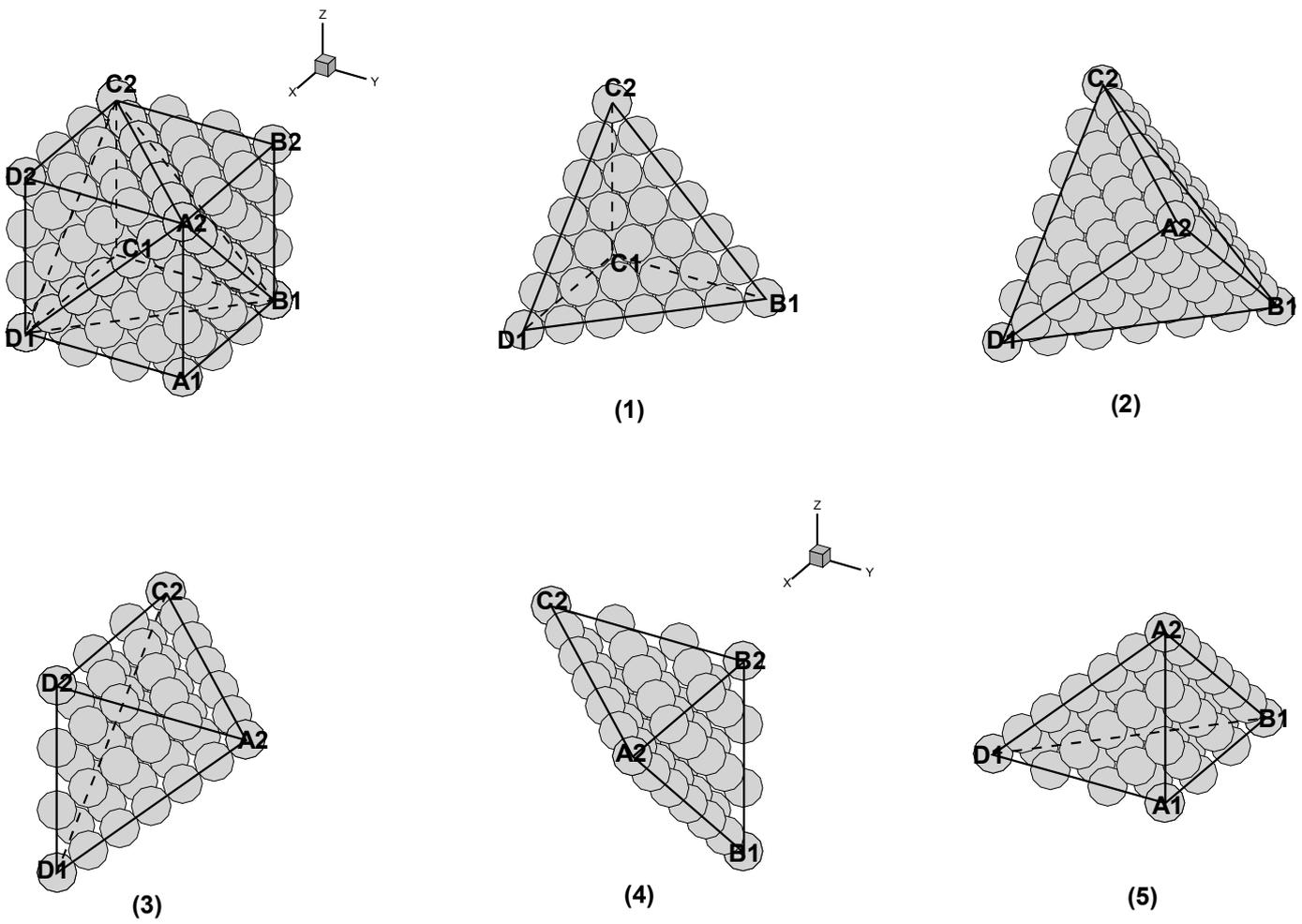

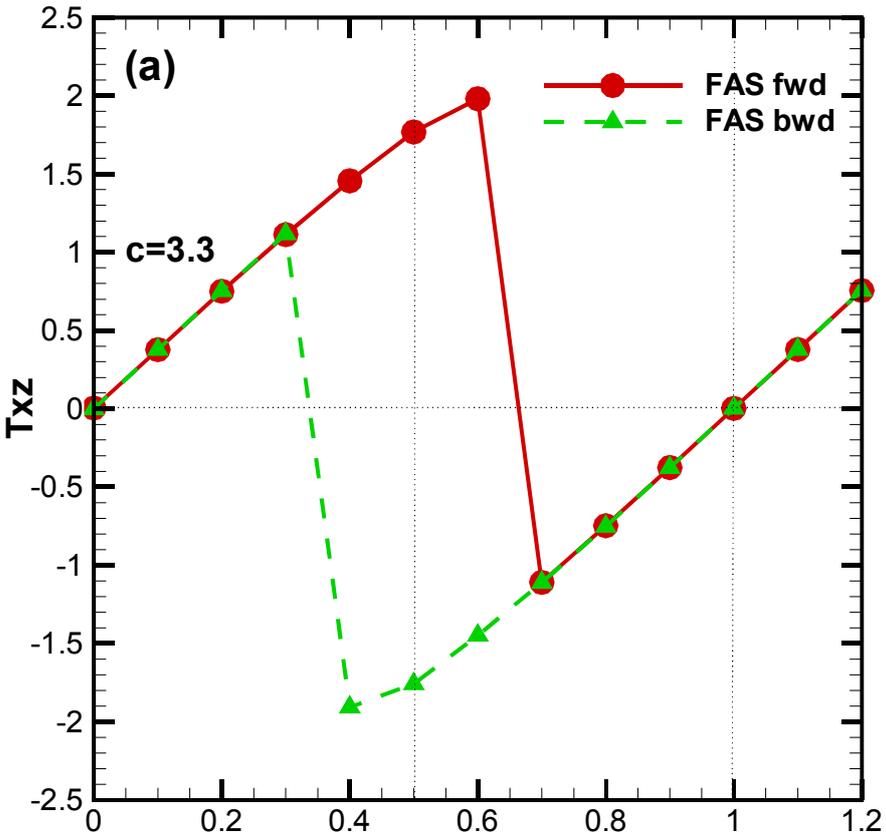

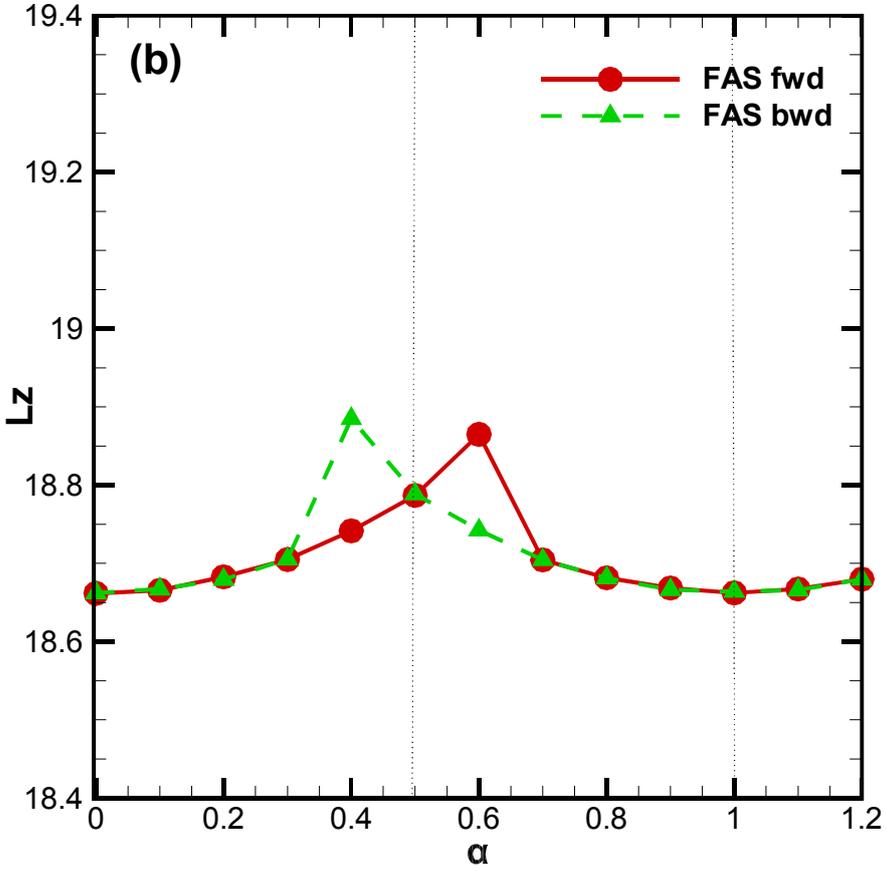

**(a)** **(b)**

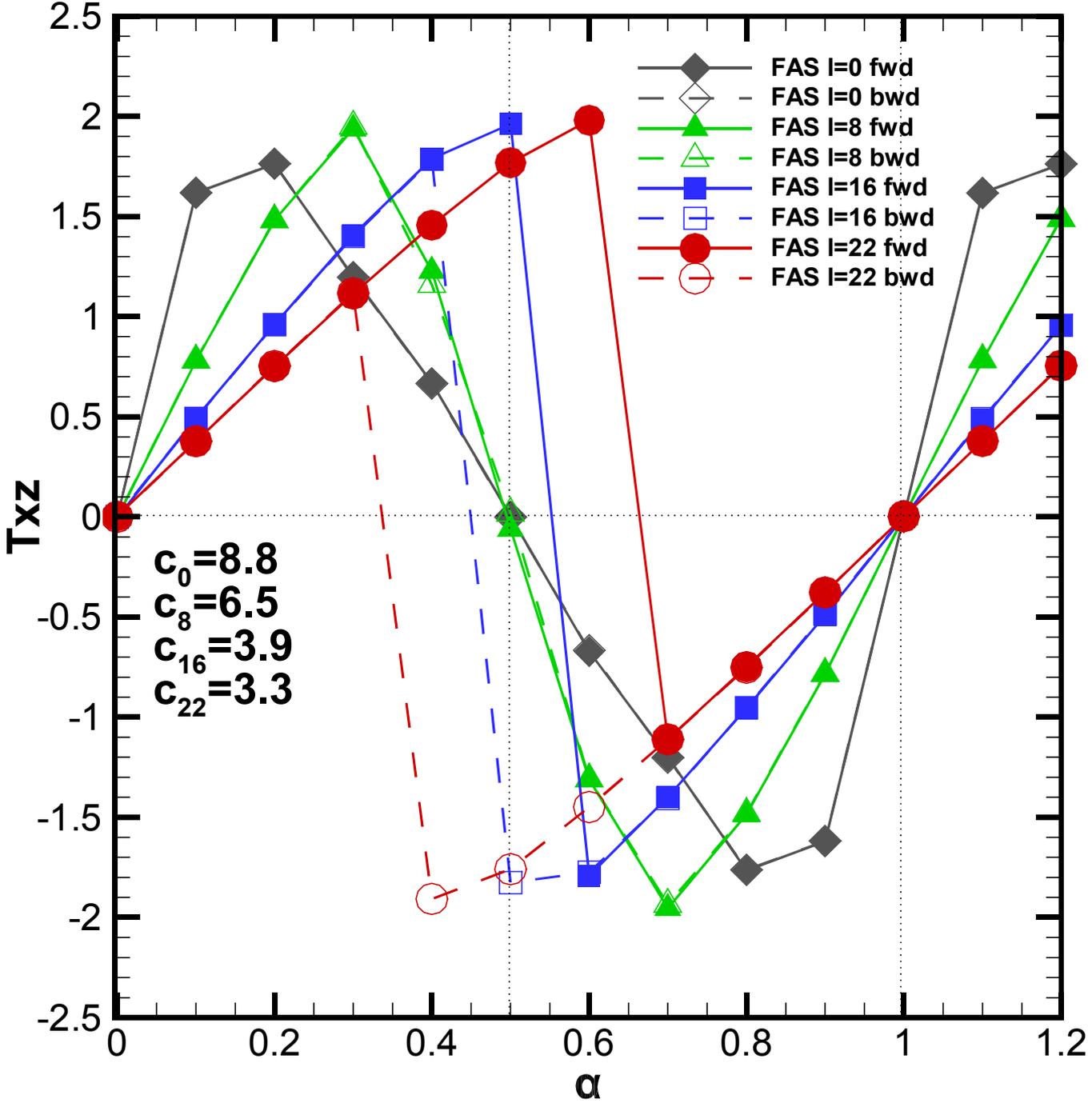

**Figure 6**
Click here to download Figure: Figure6.eps

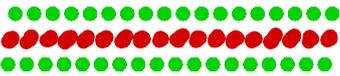 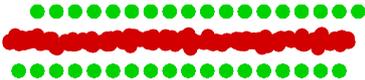 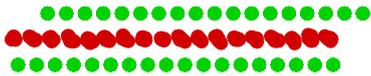

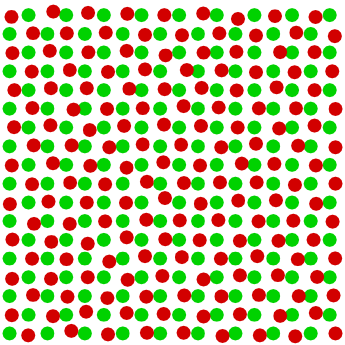 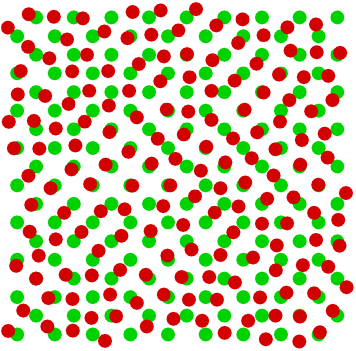 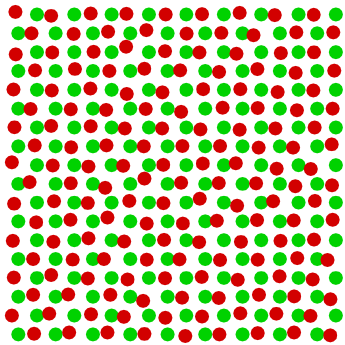

(a) (b) (c)

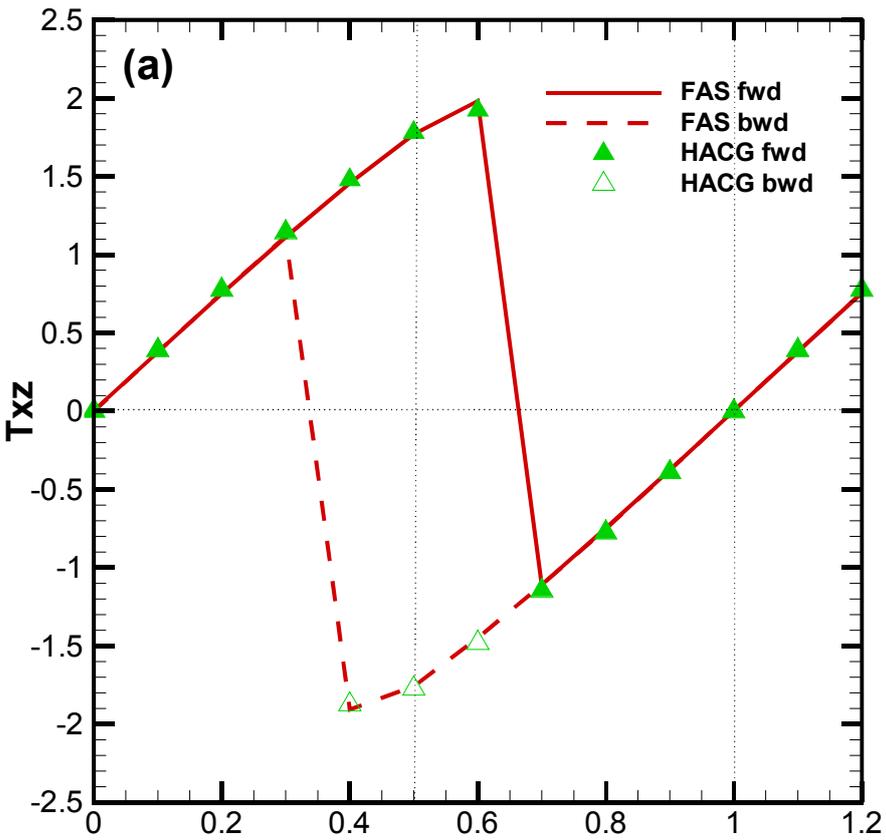

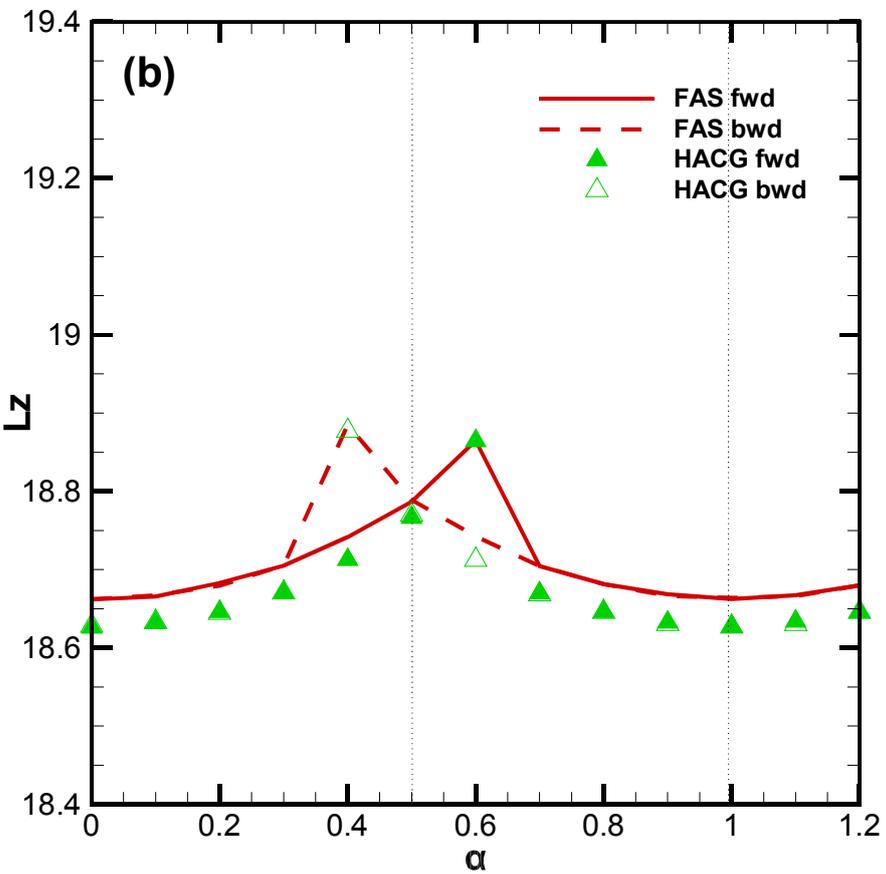

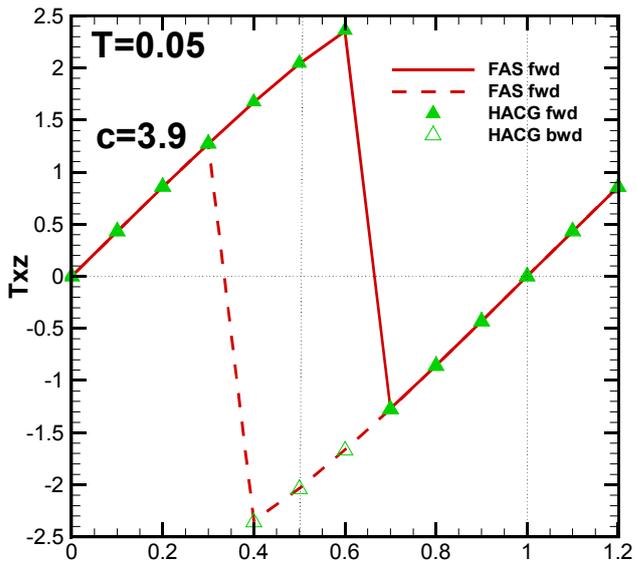

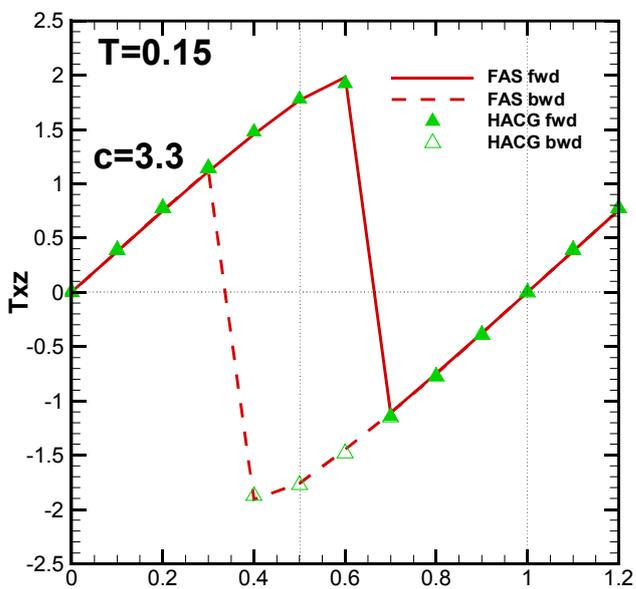

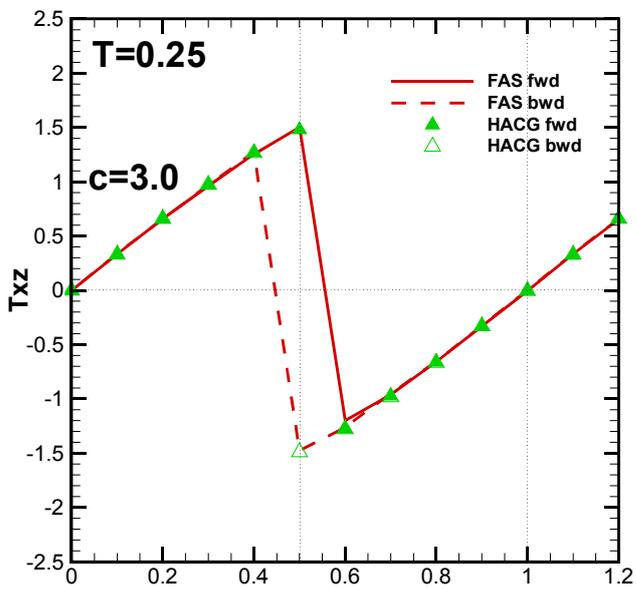

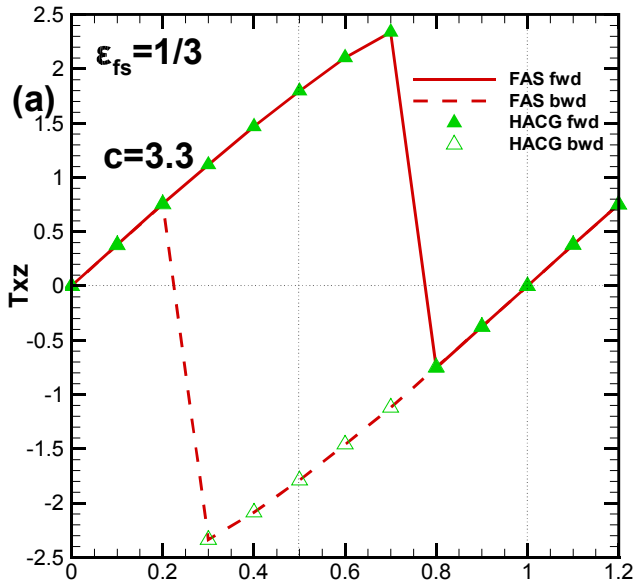

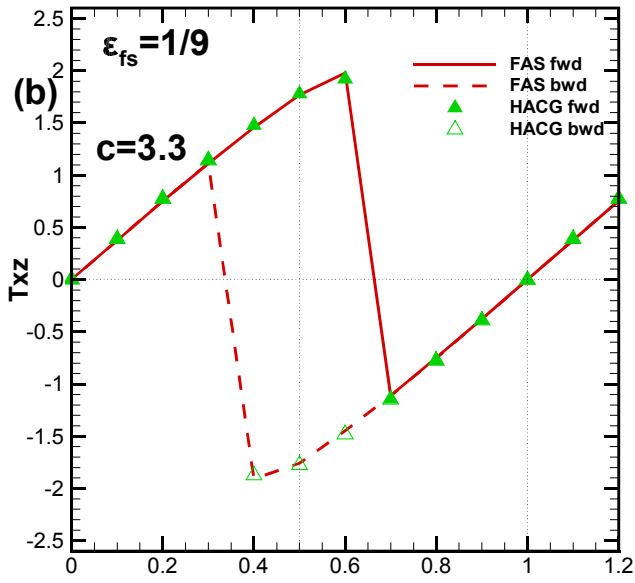

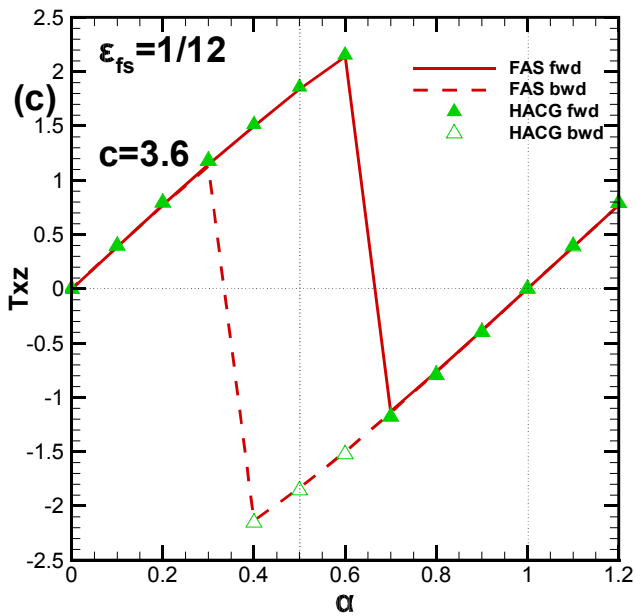

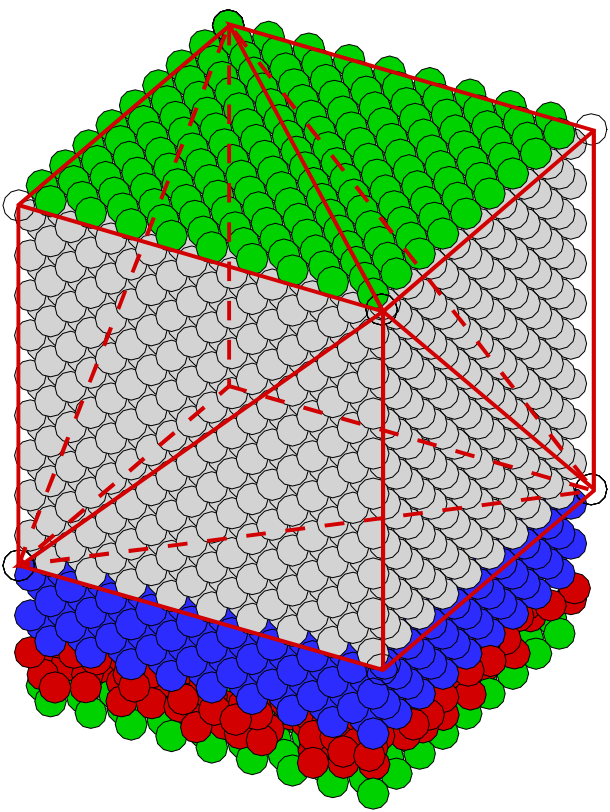
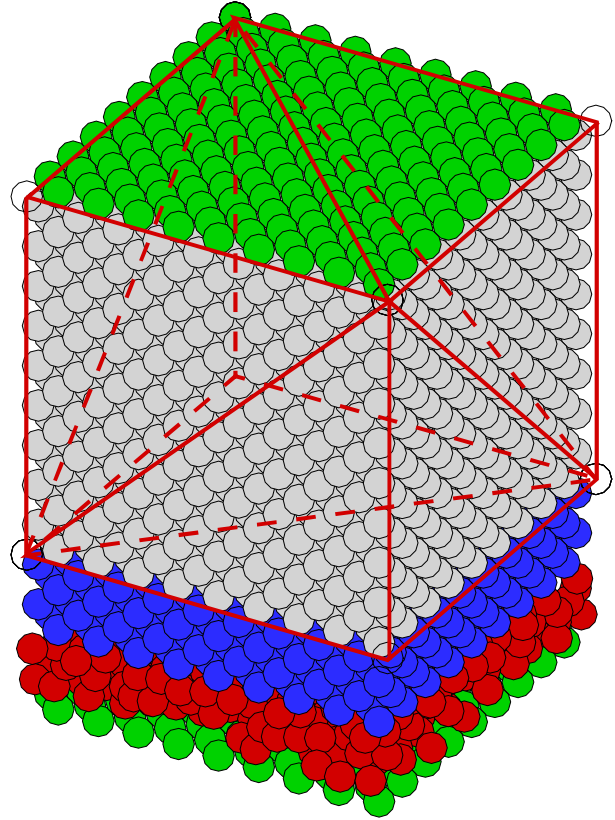

**(a)**  **(b)**

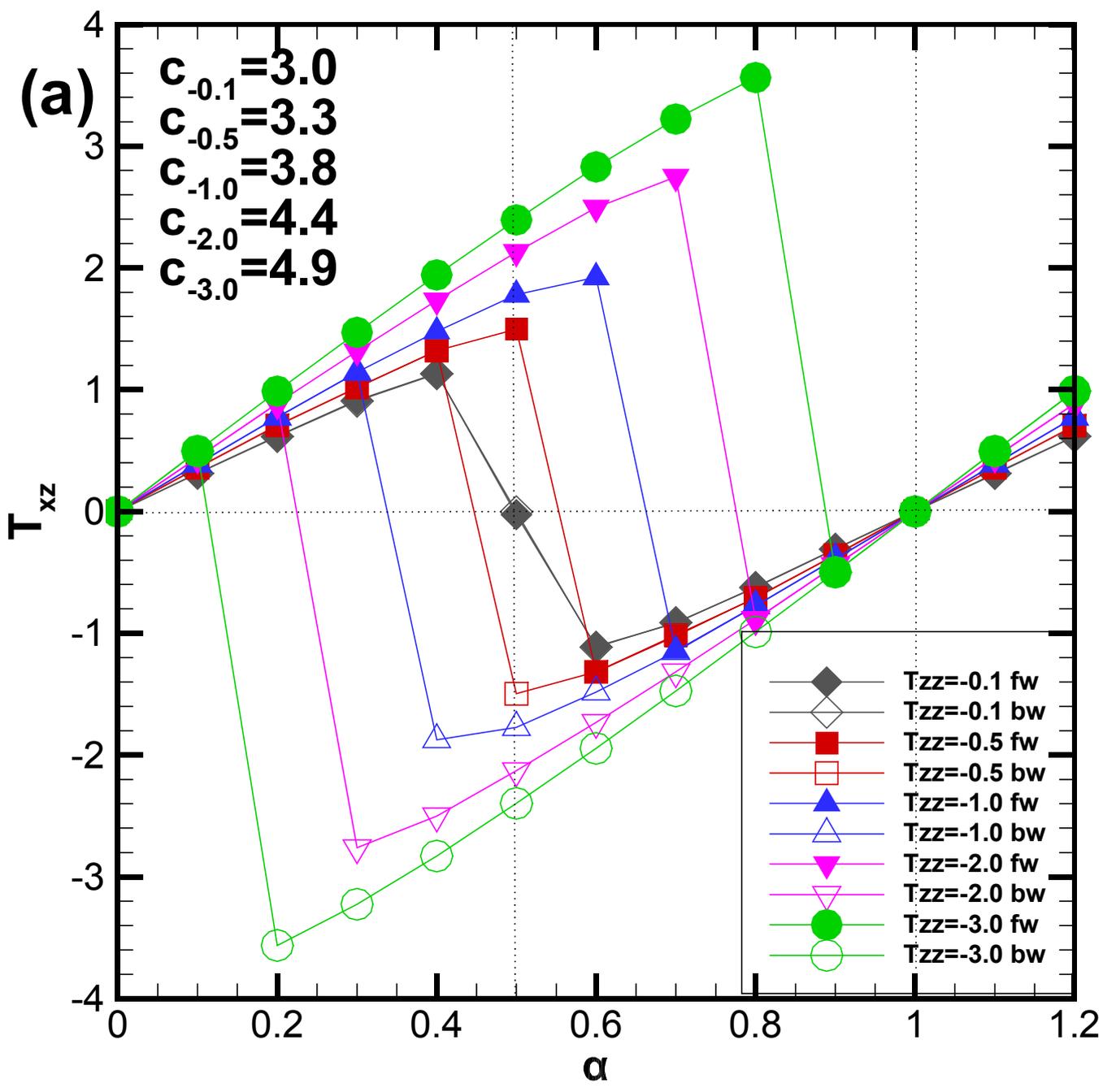

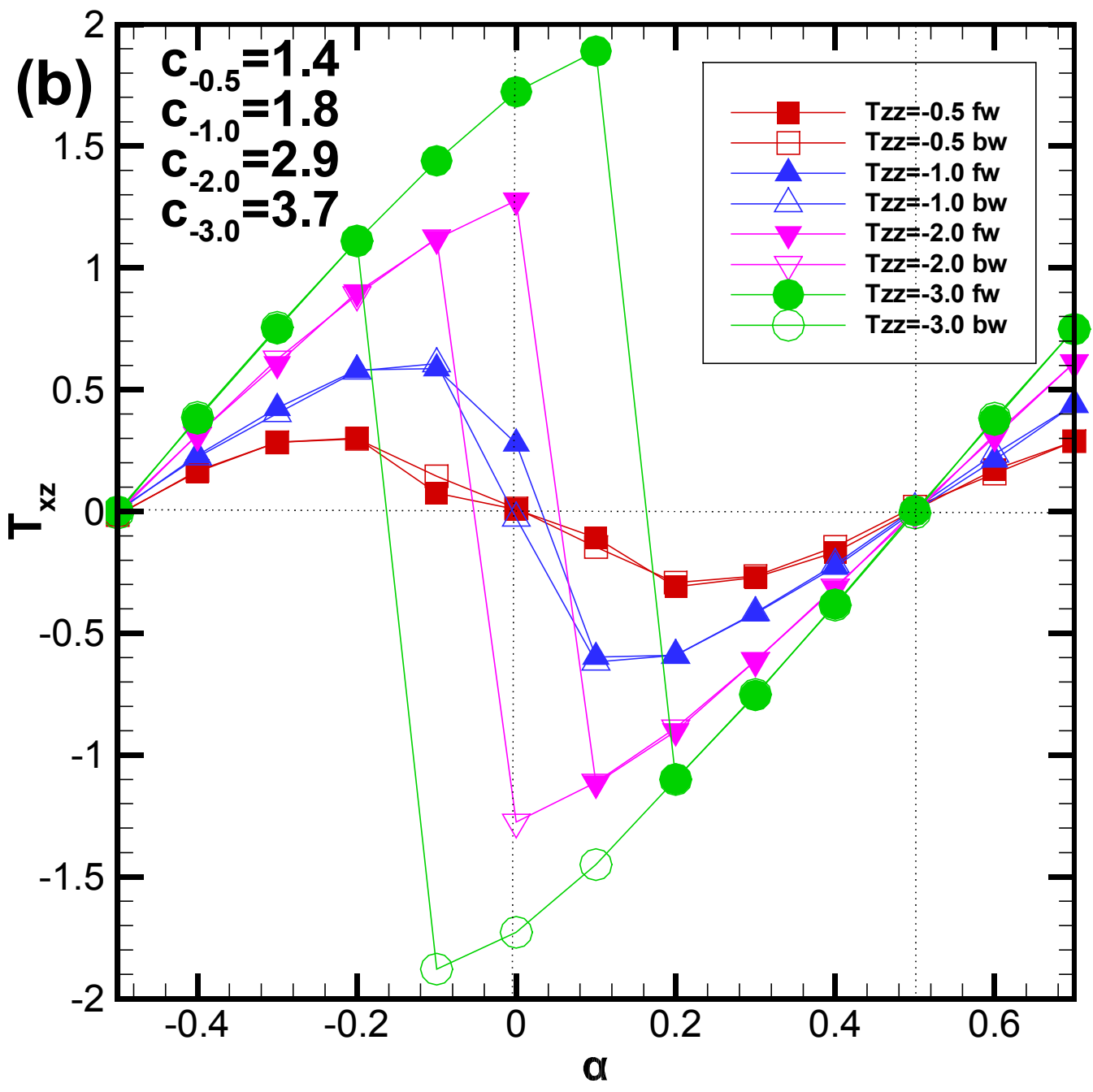

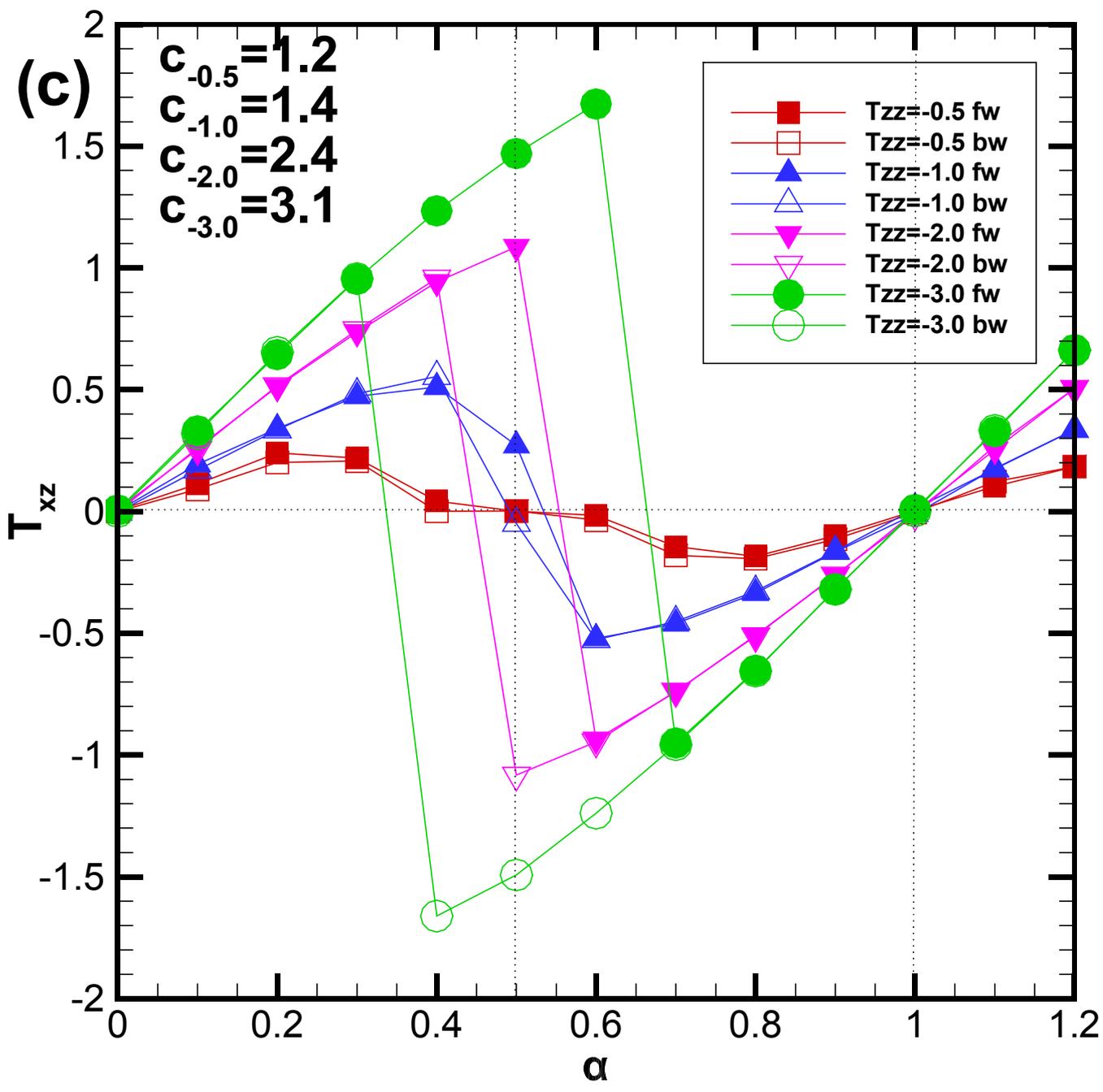

**Figure 12**
**Click here to download Figure: Figure12.eps**

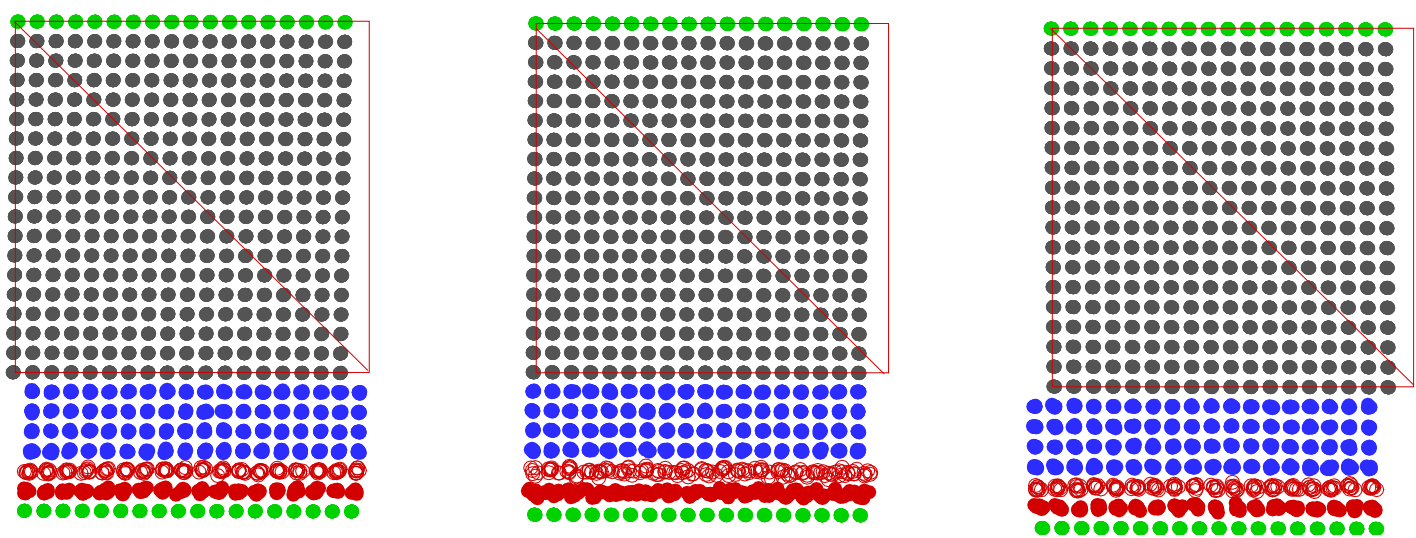

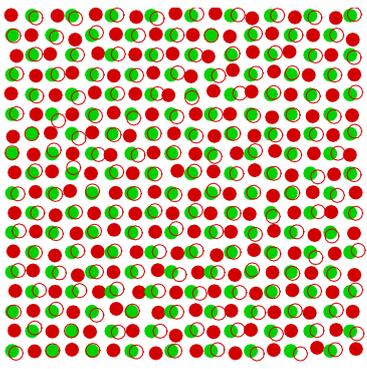
(a)

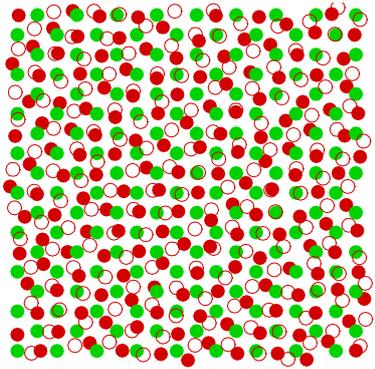
(b)

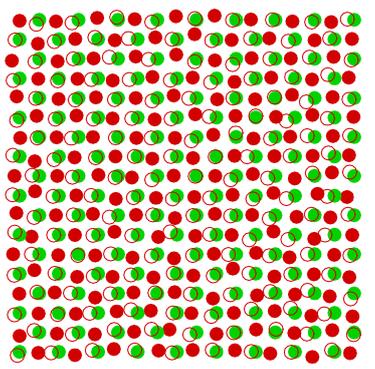
(c)

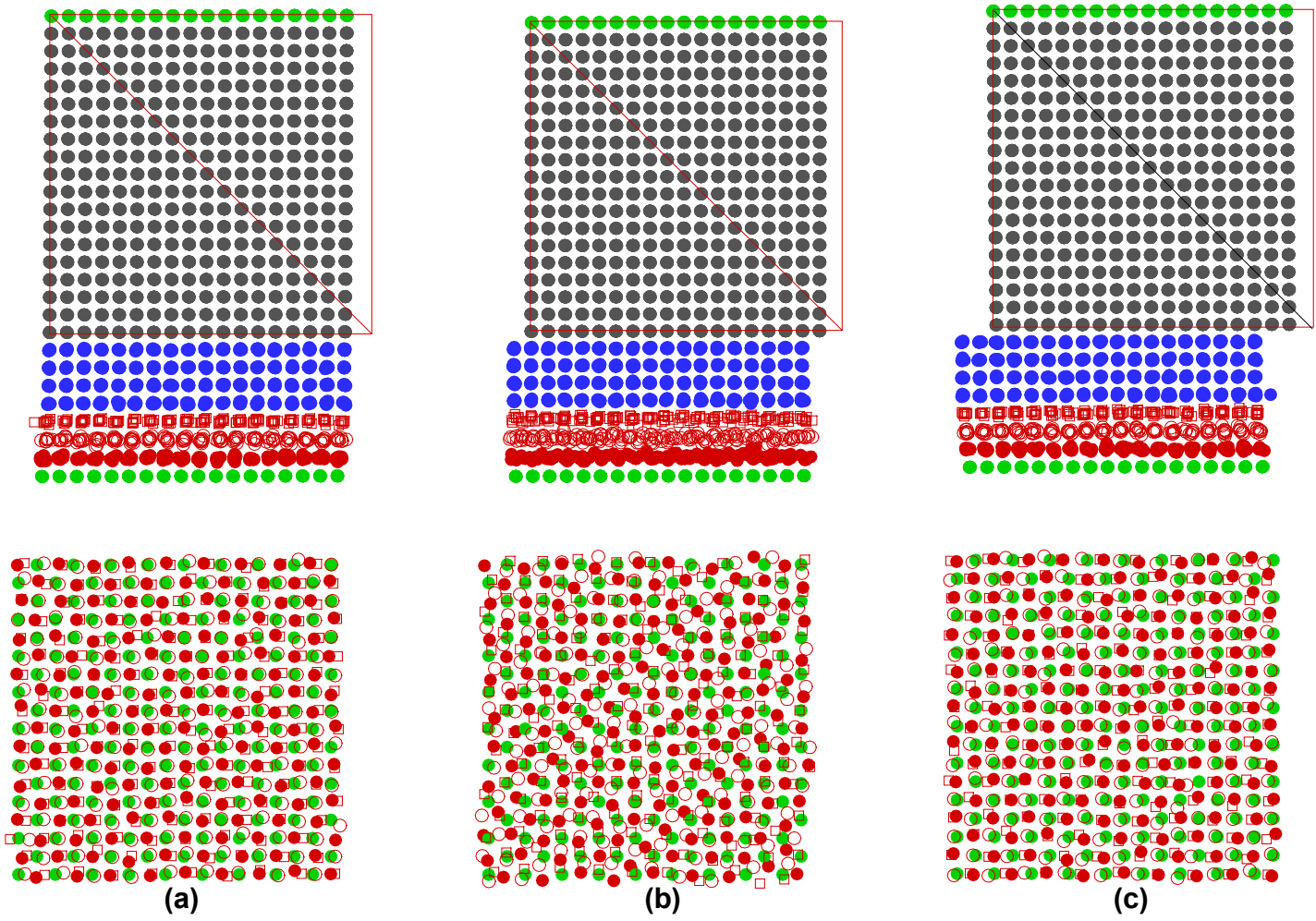

(a) (b) (c)

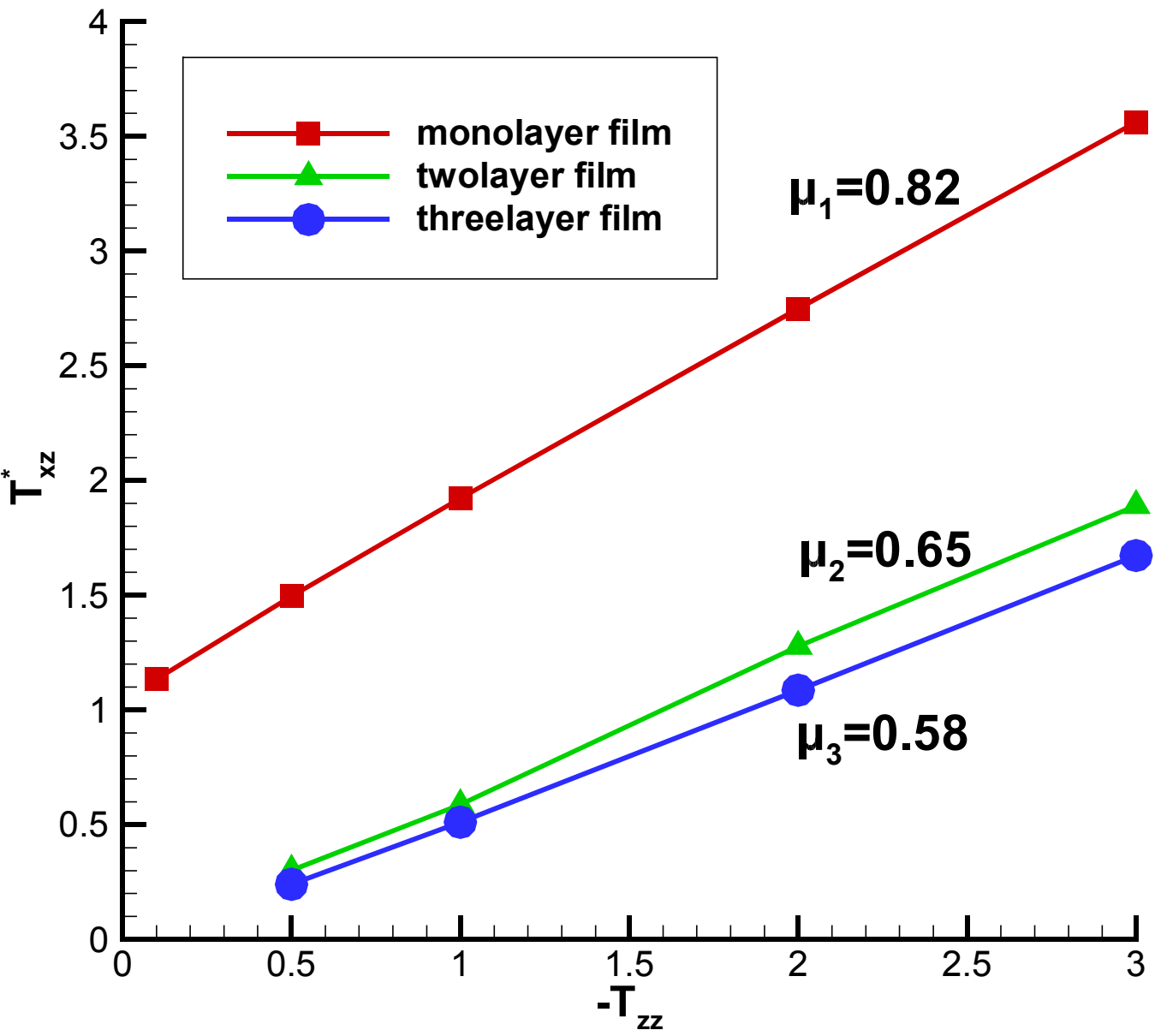